\documentclass[twocolumn]{aastex62}
\hypersetup{linkcolor=red,citecolor=blue,filecolor=cyan,urlcolor=blue}
\usepackage{graphicx,color}
\usepackage{amssymb}
\usepackage{amsmath}

\usepackage{natbib}
\usepackage{txfonts}
\usepackage{multirow}
\usepackage{array}
\usepackage{rotating}
\usepackage{sidecap}
\usepackage{hyperref}
\usepackage{epstopdf}
\usepackage{footnote}
\usepackage{tabularx}
\usepackage{booktabs}
\usepackage{longtable}
\usepackage{tabu}
\usepackage{longtable}

\newcommand{\kms}{km~s$^{-1}$}

\newcommand{\Hinode}{\textit{Hinode}}

\newcommand{\sdo}{\textit{SDO}}
\newcommand{\iris}{\textit{IRIS}}

\shortauthors{Panesar et al.}

\shorttitle{Solar Jetlets}
\shortauthors{Panesar et al.}

\begin{document}

\title{IRIS and SDO Observations of Solar Jetlets Resulting from Network-Edge Flux Cancelation}

\correspondingauthor{Navdeep K. Panesar}
\email{navdeep.k.panesar@nasa.gov}
\email{alphonse.sterling@nasa.gov}

\author[0000-0001-7620-362X]{Navdeep K. Panesar}
\altaffiliation{Also, Visiting Scholar at W. W. Hansen Experimental Physics Laboratory,\\ Stanford University, Stanford, CA 94305, USA}
\affil{NASA Marshall Space Flight Center, Huntsville, AL 35812, USA}

\author{Alphonse C. Sterling}
\affiliation{NASA Marshall Space Flight Center, Huntsville, AL 35812, USA}
\altaffiliation{alphonse.sterling@nasa.gov}

\author{Ronald L. Moore}
\affiliation{Center for Space Plasma and Aeronomic Research (CSPAR), UAH, Huntsville, AL 35805, USA}
\affiliation{NASA Marshall Space Flight Center, Huntsville, AL 35812, USA}

\author{Sanjiv K. Tiwari}
\affiliation{Lockheed Martin Solar and Astrophysics Laboratory, Org. A021S, Bldg. 252,  3251 Hanover St., Palo Alto, CA 94304, USA}
\affiliation{Bay Area Environmental Research Institute Palo Alto, Palo Alto, USA}

\author{Bart De Pontieu}
\affiliation{Lockheed Martin Solar and Astrophysics Laboratory, Org. A021S, Bldg. 252,  3251 Hanover St., Palo Alto, CA 94304, USA}
\affiliation{Rosseland Centre for Solar Physics, University of Oslo, P.O. Box 1029 Blindern, NO–0315 Oslo, Norway}
\affiliation{Institute of Theoretical Astrophysics, University of Oslo, P.O. Box 1029 Blindern, NO–0315 Oslo, Norway}

\author{Aimee A. Norton}
\affiliation{W. W. Hansen Experimental Physics Laboratory, Stanford University, Stanford, CA 94305, USA}


\begin{abstract}

Recent observations show that the buildup and triggering of minifilament eruptions that drive coronal jets result from magnetic flux cancelation at the neutral line between merging majority- and minority-polarity magnetic flux patches. We investigate the magnetic setting of ten on-disk small-scale UV/EUV jets (\textit{jetlets}, smaller than coronal X-ray jets but larger than chromospheric spicules) in a coronal hole by using \iris\ UV images and \sdo/AIA EUV images and line-of-sight magnetograms from \sdo/HMI. We observe recurring jetlets at the edges of magnetic network flux lanes in the coronal hole.  From magnetograms co-aligned with the \iris\ and AIA images, we find, clearly visible in nine cases, that the jetlets stem from sites of flux cancelation proceeding at an average rate of $\sim$1.5 $\times$ 10$^{18}$ Mx hr$^{-1}$, and show brightenings at their bases reminiscent of the base brightenings in larger-scale coronal jets. We find that jetlets happen at many locations along the edges of network lanes (not limited to the base of plumes)  with average lifetimes of 3 min and  speeds of 70\kms. The average jetlet-base width (4000 km) is three to four times smaller than for coronal jets ($\sim$18,000 km).
Based on these observations of ten obvious jetlets, and our previous observations of larger-scale coronal jets in quiet regions and coronal holes, we infer that flux cancelation is an essential process in the buildup and triggering of jetlets. Our observations suggest that network jetlet eruptions might be small-scale analogs of both larger-scale coronal jets and the still-larger-scale eruptions producing CMEs.

\end{abstract}


\section{Introduction} \label{sec:intro}

Solar jets are short-lived, collimated, transient events frequently observed in the solar atmosphere \citep{shibata92,innes11,raouafi16}. Jets of all sizes, down to and including spicules \citep{pontieu07,sterling16b}, plausibly power the heating of the non-active-region global corona \citep{moore99,pontieu11}. Coronal jets (CJs) occur in various solar environments (e.g., in active regions, quiet Sun regions and in coronal holes), and launch plasma high into the corona \citep{wangYM98,moore15,panesar16a}. 

Recent observations show that CJs are frequently driven by the eruption of a \textit{minifilament} (\citealt{sterling15}; also see \citealt{hong11,shen12,adams14}) and a jet bright point (JBP) appears under the minifilament as it erupts to drive the jet. The JBP is a miniature version of the flare arcades that grow over polarity inversion lines (PILs) in the wake of larger-scale filament eruptions that drive coronal mass ejections (CMEs). Thus, CJs are analogous to typical larger-scale solar eruptions \citep{zir89,mar73}.

Recently, we investigated the triggering mechanism of 10 quiet-region \citep{panesar16b,panesar17} and 13 coronal-hole \citep{panesar18} CJs and 
found that CJs in these regions are driven by the eruption of a minifilament, and that the minifilament magnetic field is built (minifilaments exist for periods ranging from 1.5 hr to 2 days prior to their eruption) and triggered by magnetic flux cancelation at the PIL underneath the minifilament. 

{Using \textit{Solar Dynamic Observatory} (\sdo)/Atmospheric Imaging Assembly (AIA; \citealt{lem12}), \cite{raouafi14} found jet-like features at 
flux-cancelation sites at the bases of plumes.  They named these features \textit{jetlets}, because 
they are smaller than typical CJs.  In the region we study here, we found jetlets at the base 
of plumes, but we also found identical-looking features outside of plumes too; we call all of 
these features jetlets. In this study we characterize jetlet properties, compare their properties 
to CJs, and try to determine whether they are small-scale counterparts to CJs.

We investigate in detail the magnetic setting of 10 on-disk solar UV/EUV network jetlets and examine their physical properties using \textit{Interface Region Imaging Spectrograph} (\iris; \citealt{pontieu14}) and \sdo\ data. We find that flux cancelation is the cause of most of our jetlets, and that they often occur at the edges of network lanes away from the bases of plumes. \cite{raouafi14} only reported  jetlets that happened at the base of plumes; we find that jetlets are more wide spread in network regions, not limited to only the base of plumes.

\floattable

\begin{table*}
	
	\setlength{\tabcolsep}{2.pt} 
	
	\scriptsize{
		\caption{Location and measured parameters of observed jetlets\label{tab:list}}
		\renewcommand{\arraystretch}{1.2}
		\begin{tabular}{c*{13}{c}}
			\noalign{\smallskip}\tableline\tableline \noalign{\smallskip}
			Jetlet\tablenotemark{a} &  No. of\tablenotemark{b} &  Time\tablenotemark{c}& IRIS   & Spire Length\tablenotemark{d}  & Spire Length\tablenotemark{e} & Spire Width\tablenotemark{f} & Spire Width\tablenotemark{g}  & Jetlet Speed\tablenotemark{h} & Jetlet Dur.\tablenotemark{i} &Jetlet-Base\tablenotemark{j}  & Jetlet-Base\tablenotemark{k}   & $\Phi$ Cancelation\tablenotemark{l}   \\
			
			Location   &  Jetlets  & (UT) & Coverage & IRIS (km) & AIA (km) & IRIS (km) & AIA (km) & (\kms) & (minutes) & AIA (km)& IRIS (km) &  rates 10$^{18}$ Mx hr$^{-1}$   \\
			
			\noalign{\smallskip}\hline \noalign{\smallskip}
			A   & 1 (A1) & 22:07  &   No &- &11000$\pm$500 &- &2300$\pm$500 &23$\pm$1 & 4$\pm$12s & 3200$\pm$500 &- & 1.0 \\ 
			
			B   & 2 (B1) & 13:57, &  No  &- &-\tablenotemark{m} &- &- &  -            & 2$\pm$24s &2200$\pm$500 &- & -\tablenotemark{m} & \\ 
			& (B2)  & 19:04\tablenotemark{n}& Yes& 6900$\pm$800 &- &800$\pm$70 &- &- & 2$\pm$12s & 2100$\pm$200 &1000$\pm$80 & - & \\ 
			C   & 3 (C1)& 16:40, &  No  &- &25000$\pm$2500 &- &5000$\pm$25 & 110$\pm$25 & 2$\pm$12s&2500$\pm$400 &- & 1.7  \\ 
			&  (C2) & 18:33, &  Yes &19000$\pm$1000 &-\tablenotemark{o} &1300$\pm$300 &-\tablenotemark{o} & -\tablenotemark{o}          &3$\pm$24s& 3000$\pm$300 &4200$\pm$300  & \\
			&  (C3) & 19:15  &  Yes &23000$\pm$1000 &33000$\pm$1800 &2100$\pm$200 &2500$\pm$600 & 120$\pm$50 & 4$\pm$24s&5000$\pm$800 &3000$\pm$200 & \\  
			
			D   & 1 (D1) & 19:15  &  Yes &11000$\pm$800 &18000$\pm$1500 &4000$\pm$300 &1500$\pm$400 & 50$\pm$20  & 3$\pm$12s&4300$\pm$200 &6000$\pm$250 & 2.6  \\ 
			
			E   & 3 (E1) & 21:16, &  Yes &16500$\pm$300 &33000$\pm$3500 &7000$\pm$100 &2200$\pm$150 & 70$\pm$30  &3$\pm$12s& 10,000$\pm$1000 &8000$\pm$100 & 0.6  \\

			&  (E2) & 22:37, &  No  &- &32000$\pm$2000 &- &2000$\pm$150 & 60$\pm$10 &4$\pm$24s&5000$\pm$300 &- & \\
			& (E3)  & 23:14  &  No  & -&33600$\pm$1200 &- &7000$\pm$150 & 50$\pm$10 &5$\pm$12s&6500$\pm$200 &- & \\

			\noalign{\smallskip}\tableline\tableline \noalign{\smallskip}
			average$\pm$1$\sigma$$_{ave}$ & & & & 16000$\pm$6000&27000$\pm$8000&3000$\pm$2500&3200$\pm$2000&70$\pm$30&3$\pm$1&4000$\pm$2000&4000$\pm$2500& 1.5 &\\
	\hline
			
		\end{tabular}
		
		\tablenotetext{a}{ Jetlet locations in Figure \ref{fig1}c.}
		\tablenotetext{b}{ Total number of jetlets from the same neutral line.} 
		\tablenotetext{c}{ Approximate time of brightening at the base of jetlets in AIA 171 \AA\ images. }
		\tablenotetext{d,e}{~~~~Maximum length of the spire from base to the visible tip of the spire.}
		
		\tablenotetext{f,g}{~~~ Width measured in the middle of the spire using AIA 171\AA\ and \iris\ Si IV SJI. }
		\tablenotetext{h}{ Plane-of-sky speed along the jetlet spire. Speeds and uncertainties are measured from  AIA 171\AA\ time-distance maps.}
		\tablenotetext{i}{ Duration of jetlet spire in AIA 171\AA\ images. }
		\tablenotetext{j, k}  {~~~ Cross-sectional width of the jetlet-base during the jetlet onset in  AIA 171\AA\ and \iris\ Si IV SJI.}
		\tablenotetext{l}{ Average flux cancelation rates from 1-2 hours before the jetlet to 0-1 hours after the jetlet.}
		\tablenotetext{m}{ Jetlet is barely visible in AIA images. HMI magnetograms do not show minority-polarity flux at the base of the jetlet.}
		\tablenotetext{n}{ There are small-scale jetlets around that time in \iris\ images.}
		\tablenotetext{o}{ Spire is too faint in AIA images to estimate its length, width, and speed.}
		
	}
	
\end{table*}

\begin{figure}
	\centering
	\includegraphics[width=\linewidth]{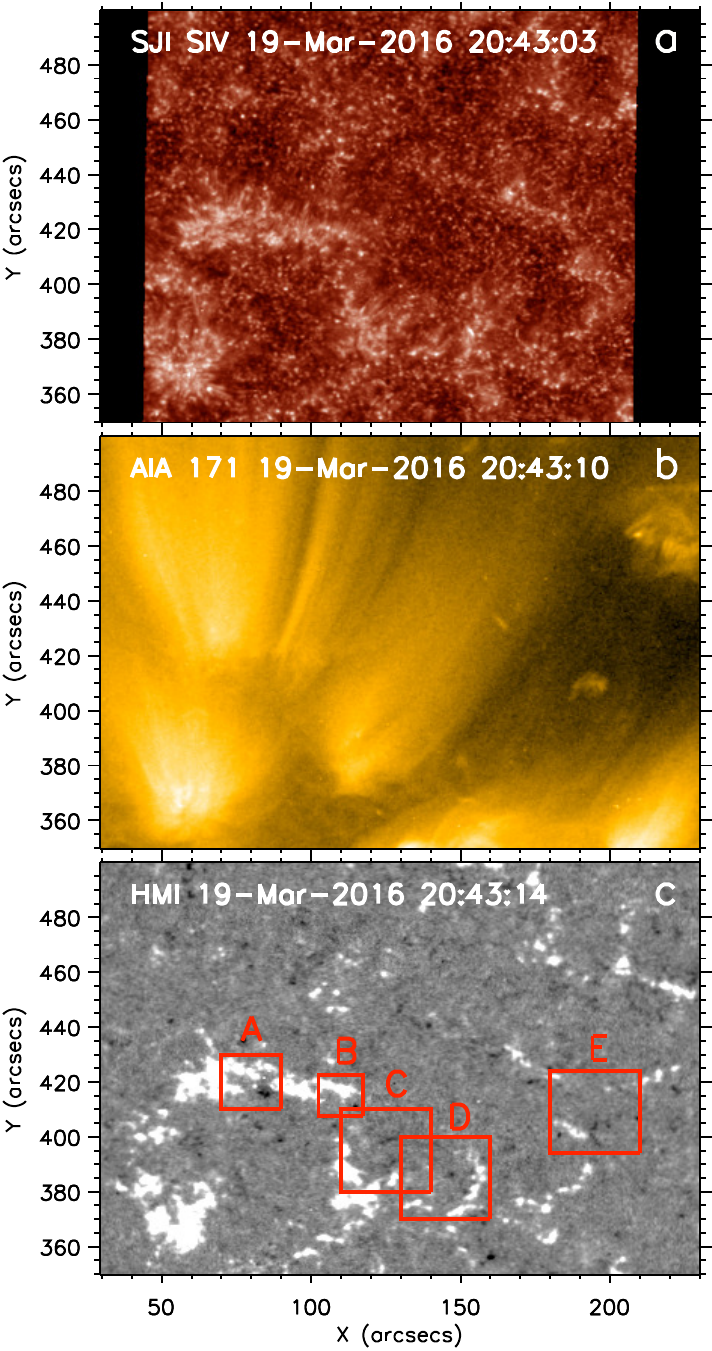}
	\caption{ Locations of the ten jetlets of Table \ref{tab:list}: Panel (a) shows an \iris\ Si IV SJI of the coronal hole region. Panels (b) and (c) show  an AIA 171 \AA\ image and an HMI magnetogram of the same region. The red boxes show the FOVs analyzed in detail and multiple jetlets appear within this FOV.} \label{fig1}
\end{figure} 

\begin{figure*}
	\centering
	\includegraphics[width=0.8\linewidth]{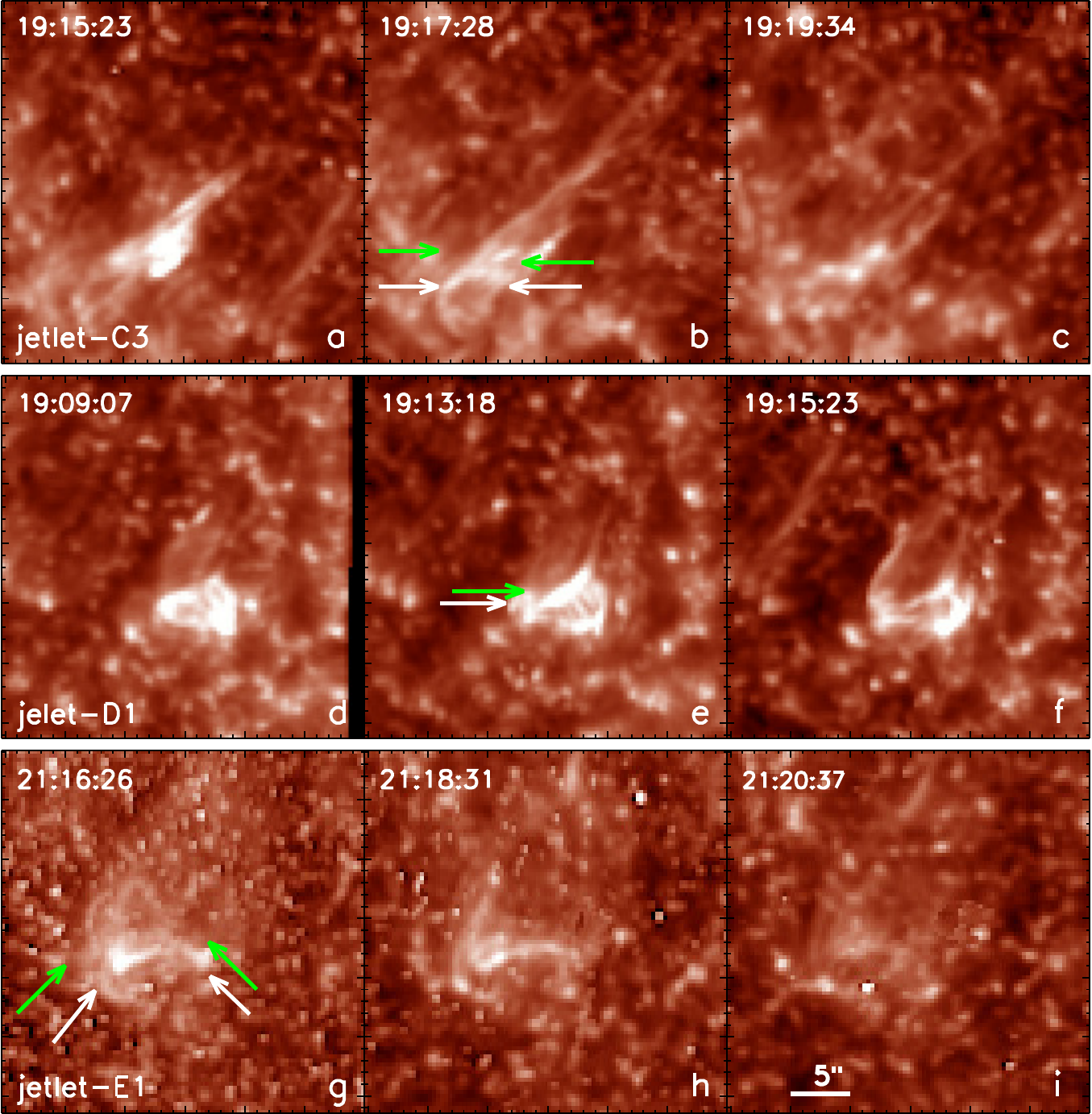}
	\caption{\iris\ Si IV SJIs of the evolution of three jetlets. Panels (a)-(c) show a jetlet (C3) from location C of Figure \ref{fig1}. Panels (d)-(e) show a jetlet  (D1) from location D of Figure \ref{fig1}. Panels (g)-(i) show a jetlet (E1) from location E of Figure \ref{fig1}. The white arrows point to the Si IV brightenings at the base of 171 \AA\ flare loops. The green arrows point to the locations of the feet of the bright loops in 171 \AA\ images. An animation of this figure is available. The temporal cadence of each frame is $\sim$2min.} \label{fig2}
\end{figure*} 
\section{INSTRUMENTATION AND DATA}\label{data} 

\iris\ provides simultaneous images (slit-jaw) and spectra of the solar 
atmosphere with high spatial resolution of 0\arcsec.16 pixel$^{-1}$  and cadence as high as 1.5s in four different passbands (C II 1330, Si IV 1400, Mg II k 2796, and Mg II wing 2830 \AA; \citealt{pontieu14}). For our investigation, we used C II and Si IV slit-jaw images (SJIs) having a cadence of 2~min. 

We also used 171 \AA\ extreme ultraviolet (EUV) images from \sdo/AIA\ to view the coronal-temperature plasma in the jetlets.  To study the magnetic field evolution of the jetlet base region, we employed line-of-sight magnetograms from the \sdo/Helioseismic and Magnetic Imager (HMI; \citealt{schou12}). 

For our analysis, we randomly selected an on-disk coronal hole network region covered by \iris\ on 19-March-2016\footnote{\url{http://www.lmsal.com/hek/hcr?cmd=view-event\&event-id=ivo\%3A\%2F\%2Fsot.lmsal.com\%2FVOEvent\%23VOEvent_IRIS_20160319_181933_3601112078_2016-03-19T18\%3A19\%3A332016-03-19T18\%3A19\%3A33.xml}} during 18:19--21:48UT. During the 3.5 hours of \iris\ coverage, we find five jetlets  at five different locations/PILs. We intentionally avoided CJs for this study (e.g.~\citealt{panesar16b}). AIA images show some plumes in the coronal hole. We find three jetlet locations/PILs at the base of plumes \citep{raouafi14,ellis18} and two jetlet locations/PILs away from plumes (Figure \ref{fig1}). None of these jetlets were covered by the \iris\  spectral slit.

Simultaneously, we study the same network region using AIA 171 \AA\ images for 24 hours centered on the \iris\ coverage-period to see if there are more jetlets from the same PILs. We find five more jetlets  within the \iris\ field-of-view  (FOV; Figure \ref{fig1}) but outside the \iris\ observation time.  All ten jetlets and their measured parameters are listed in  Table \ref{tab:list}. Out of the ten jetlets seen in AIA, only five of them  (Table \ref{tab:list}) were observed by \iris.

\begin{figure*}
	\centering
	\includegraphics[width=0.8\linewidth]{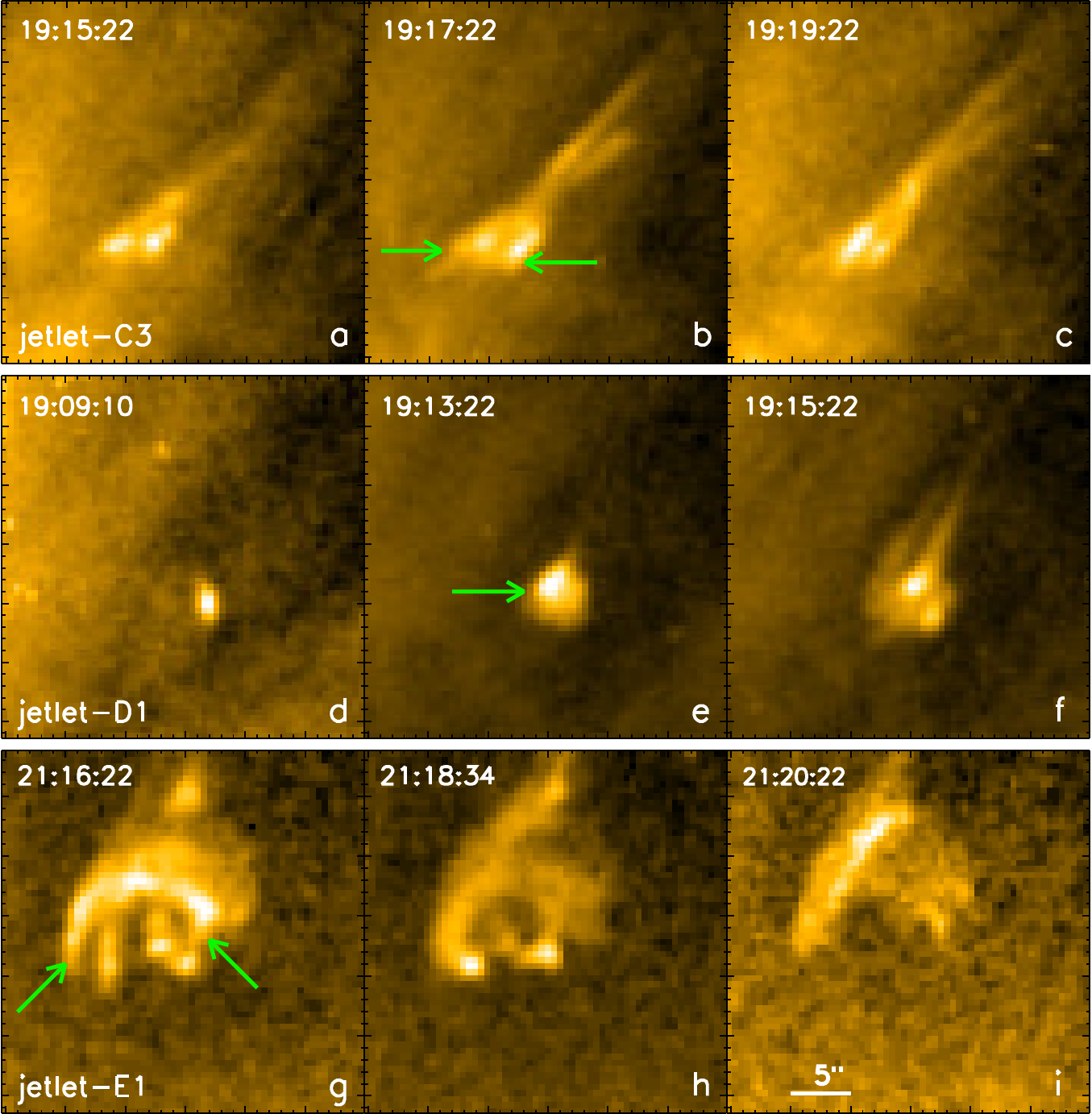}
	\caption{AIA 171 \AA\ images of the evolution of the three jetlets of Figure \ref{fig2}. The format is similar to Figure \ref{fig2}. The green arrows point to the feet of bright loops. An animation of this figure is available. The temporal cadence of each frame is $\sim$1min.} \label{fig3}
\end{figure*} 

\vspace{0.1cm}
\section{Results}

\subsection{Overview}\label{over}

 Figure \ref{fig1} shows the coronal hole region observed by \iris, \sdo\ and HMI. Jetlet locations are marked by red boxes in Figure \ref{fig1}c. In this coronal hole, positive flux is in the majority, and nine of the ten jetlets originate from the PIL  between a majority-polarity network flux lane and a merging minority-polarity flux patch.

 Jetlets from locations A, B and C occur at the base of plumes, but the other two jetlet locations (D and E) are away from plumes. More generally, AIA 171 \AA\ images show that jetlets are very common in network regions. They occur at the edges of the magnetic network lanes, and are rather frequent; we find $\sim$40 jetlets in 24 hours within the AIA FOV of Figure \ref{fig1}b. For the present study, we examine only those network-edge locations at which at least one jetlet was caught in \iris\ images. Next, we present jetlets C3, D1 and E1 in detail.

\subsection{Jetlets from Region C}\label{res1}
Figures \ref{fig2}(a-c) and \ref{fig3}(a-c) show jetlet C3. Figures \ref{fig4}a,b display the photospheric magnetic field of the jetlet-base region. The three jetlets C1-C3 were homologous, in that they originated from the same PIL and had similar structure \citep{dodson77}. During jetlet C3's onset, at 19:15 UT, we observe brightenings at its base (Figure \ref{fig2}b). Concurrently, brightenings also appear in AIA 171 \AA\ images (Figure \ref{fig3}b, MOVIE2a). These base brightenings appear to include a miniature version of a JBP. After the start of base brightening, the spire extends outward with an average speed of 120$\pm$20 \kms.

The \iris\ and AIA movies show possible twisting motion in the jetlet spire over 19:17 UT-19:21 UT (MOVIE2a; Figures \ref{fig2}b,c and \ref{fig3}b,c). Such twisting motion perhaps resulted from \textit{external reconnection} \citep{sterling15} of a miniature erupting flux rope as inferred for larger coronal  jets \citep{moore15}. The earlier two jetlets from this region do not show twisting motion.

HMI magnetograms (Figures \ref{fig4}a,b) show that the jetlets originate from the PIL between a majority-polarity (positive) network flux lane and a merging minority-polarity (negative) flux patch (MOVIE3a). Figure \ref{fig4}c shows the  negative-flux plot of the jetlet-base region over four hours. We only measured the minority-polarity flux patch (negative) because it is well isolated within the box (Figure \ref{fig4}a). The negative flux decreases, which is a clear indication of flux cancelation at the base of the jetlet. We surmise that the continuous flux cancelation over $\sim$6 hours prepares and eventually triggers each of these three sequential eruptions. After the first jetlet, a significant amount of flux still remains, and that flux continues to cancel before the second homologous eruption and further cancelation apparently prepares and triggers a third  homologous eruption (the negative flux bump at 19:17 UT is from coalescence of weak flux grains unrelated to the jetlet). We estimate the average rate of flux decrease using the best-fit line in Figure \ref{fig4}c and find it to be $\sim$1.7 $\times$ 10$^{18}$ Mx hr$^{-1}$. Miniflaments in sequential CJs have also been observed to erupt and reform/reappear at the same PIL due to flux cancelation \citep{panesar17}.

\subsection{The Jetlet from Region D}\label{res2}

Figures \ref{fig2}(d-f) and \ref{fig3}(d-f) show the single jetlet in region D. The photospheric magnetic field in region D is displayed in Figures \ref{fig4}d and \ref{fig4}e. Base brightening starts at 19:09 UT (Figures \ref{fig2}d and \ref{fig3}d). At $\sim$19:12 UT the spire extends outward with an average speed of 50$\pm$20 \kms, and the total duration of the spire is 3 minutes. The Figure \ref{fig3}e green arrow points to the AIA 171\AA\ brightening that appeared at the PIL during the onset. In the Si IV SJIs, the base brightenings appear lower down (Figure \ref{fig2}e) in comparison to the base brightenings in the AIA 171 \AA\ images (this is the case in all ten jetlets). The Si~IV images show transition-region-temperature plasma whereas the 171 \AA\ images show relatively hotter coronal plasma.

There is an emergence of a bipole, at $\sim$18:52 UT, next to the majority-polarity (positive) network flux lane (MOVIE3b). One foot (negative polarity) of the newly-emerged bipole starts merging and canceling with the neighboring majority-polarity network flux lane (Figures \ref{fig4}d and \ref{fig4}e). The flux cancelation between the foot of the newly-emerged bipole and pre-existing flux results in the jetlet at 19:09 UT. This is analogous to  CJs, where a pre-jet minifilament forms and erupts due to flux cancelation between one foot of a  newly-emerged bipole and a pre-existing majority-polarity flux patch \citep[e.g.][]{panesar17}.

Figure \ref{fig4}f shows a minority-polarity flux-versus-time plot for region D. First there is an increase in the negative flux, due to the flux emergence ($\sim$20 minutes) before the jetlet onset. Later, at 19:00 UT, negative flux starts to decrease; the flux cancelation triggers the jetlet eruption. The flux cancels with an average rate of $\sim$2.6 $\times$ 10$^{18}$ Mx hr$^{-1}$. 

\begin{figure*}
	\centering
	\includegraphics[width=0.8\linewidth]{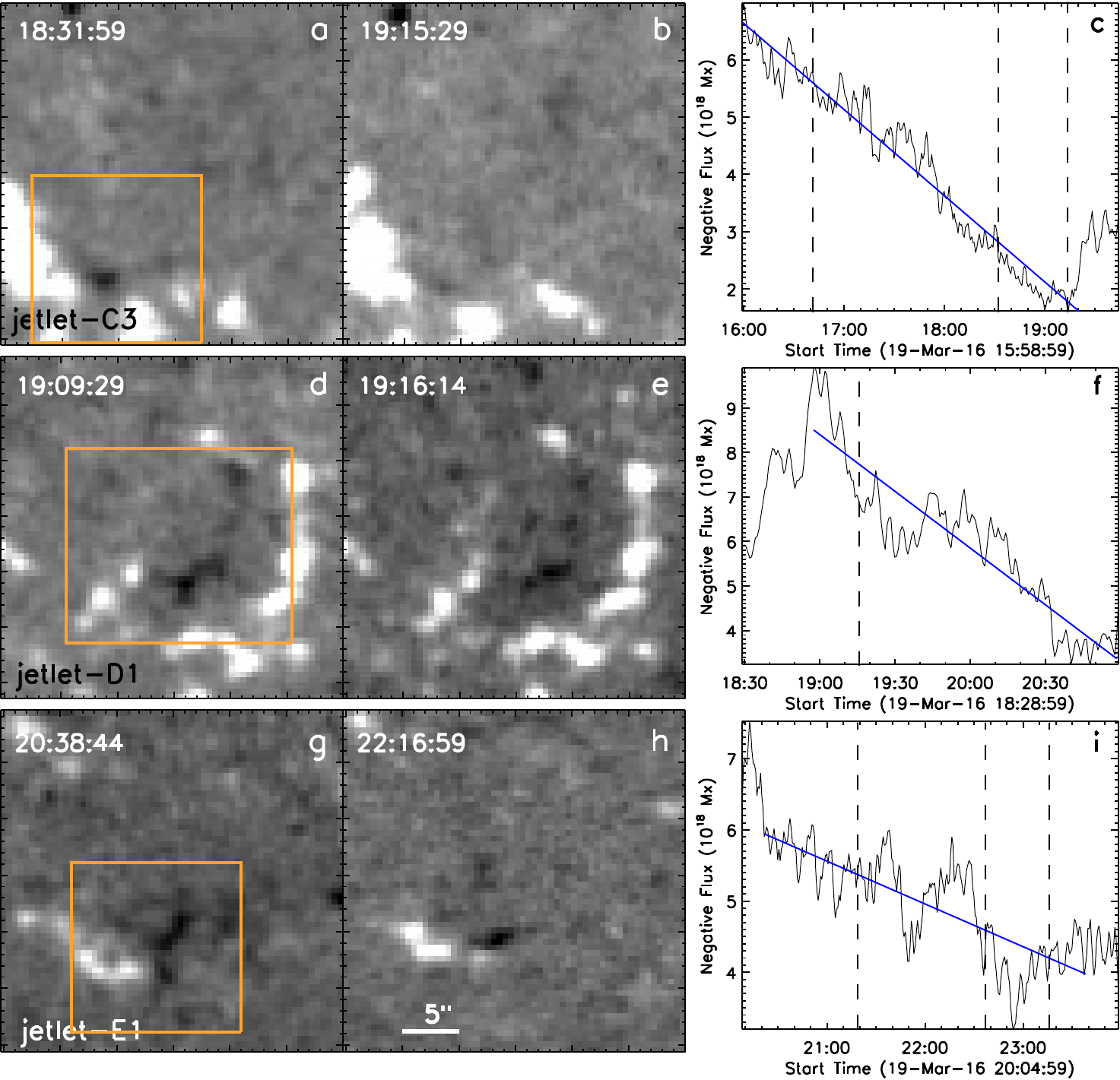} 
	\caption{HMI magnetograms of the three jetlet locations. Panels (a)-(b), (d)-(e), and (g)-(i) show  the magnetic field  near the base of jetlet C3, D1, and E1, respectively. The orange boxes in Figures \ref{fig4}a, \ref{fig4}d, and \ref{fig4}g show the regions measured for the magnetic flux time plots in Figures \ref{fig4}c, \ref{fig4}f, and \ref{fig4}i, respectively (negative flux in each case). The blue line in Figures \ref{fig4}c, \ref{fig4}f, and \ref{fig4}i is a least-square fit from before to after the jetlet. The dashed lines show jetlet onset times.  
		An animation of the a-b, d-e, and g-h panels is available. The temporal cadence of each frame is $\sim$45 seconds.}\label{fig4}
\end{figure*}

\subsection{Jetlets from Region E}\label{res3}

We show our third example jetlet, E1, in Figures \ref{fig2}(g-i) and Figures \ref{fig3}(g-i). We observed three (homologous) jetlets within a period of two hours from the same PIL.  During the onset of jetlet E1, we observe brightenings of base loops (Figures \ref{fig2} and \ref{fig3}), which link opposite-polarity flux patches (Figure \ref{fig4}). 

Figures  \ref{fig4}(g,h) show the magnetic field before and after the jetlet onset. There is a decrease in negative flux in the jetlet-base region (Figure \ref{fig4}i), which indicates that flux cancelation is the trigger of sequential/homologous jetlets. The HMI movie (MOVIE3c) shows that  opposite-polarity flux patches approach towards the PIL, and apparently flux cancelation between them triggers the first jetlet (E1) at 21:16 UT (Figure \ref{fig4}i). The flux continues to cancel after the first jetlet and continues until the minority-polarity flux patch is almost all gone (MOVIE3c). The continuous flux cancelation leads to the second (E2) and third (E3) homologous jetlets at 22:37 UT and 23:14 UT, respectively. Eventually, the minority-polarity patch is nearly gone and jetting stops.  The average flux cancelation rate is $\sim$0.6 $\times$ 10$^{18}$ Mx hr$^{-1}$.

\section{Discussion}

We have examined the magnetic environment of ten on-disk jetlets in UV/EUV in a coronal hole network region using \iris\ and \sdo\ data, and find that jetlets have many similarities with typical CJs. Recent observations of CJs  \citep[e.g.][]{huang12,shen12,panesar16b,sterling17,panesar18} show that flux cancelation is usually the trigger of CJs. Often continuous flux cancelation leads to recurrent/homologous jets \citep{chandra15,sterling16,panesar17}. CJs show base brightenings during the onset and the spire is frequently seen to extend along a twisting magnetic field \citep[e.g.][]{schmieder13,moore15,panesar16a}. Our observations show that jetlets share these properties of CJs, and hence are plausibly scaled-down versions of CJs. Here we summarize our findings:

\textbf{Flux cancelation:} We find that all ten of our jetlets occur at the edges of magnetic network flux lanes, all but one of them at a site of apparent magnetic flux cancelation. The continuous flux cancelation builds and triggers the homologous jetlet eruptions from region B, C and E. The one exception (out of ten jetlets) is jetlet B1, where HMI magnetograms do not show a minority-polarity flux patch at the jetlet base in region B. There are two possibilities in this case: either there is no minority-polarity flux present in this region, or minority-polarity flux is there but too weak to be detected by HMI. We tried increasing the HMI magnetogram sensitivity using a post-launch-improved estimate
for the instrumental point-spread function, as described in \cite{couvidat16}, but we  
were still unable to detect minority polarity in at the jetlet base in region~B. 

The average flux cancelation rate for our nine jetlets having obvious cancelation is  $\sim$1.5 $\times$ 10$^{18}$ Mx hr$^{-1}$, which is similar to that for CJs in quiet regions ($\sim$1.5 $\times$ 10$^{18}$ Mx hr$^{-1}$; \citealt{panesar16b}) and coronal holes ($\sim$0.6 $\times$ 10$^{18}$ Mx hr$^{-1}$; \citealt{panesar18}). Whereas active regions CJs have higher flux cancelation rate $\sim$1.5 $\times$ 10$^{19}$ Mx hr$^{-1}$ \citep{sterling17}.

\textbf{Base-brightenings:} During the eruption onset, we observe brightenings at the base of each of the jetlet. These brightenings occur at the PIL where opposite-polarity flux patches cancel. Jetlet-base brightenings include brightenings that might be miniatures of the JBP that is seen to occur at the flux cancelation PIL in CJs during minifilament-eruption onset.

\textbf{Twist in a jetlet spire:} In one jetlet (C3), we observe possible twisting motion in the jetlet spire, which could be the result of a miniature erupting flux rope having \textit{external reconnection} with the ambient open field. We do not see a `microfilament' of the size of the jetlet base \citep{sterling16b} in the observations. If jetlets do have `microfilament' flux ropes at their PILs before the eruption, then they plausibly work in the same way as CJs and CMEs. They may be visible in 
	higher-spatial-resolution data from current/future instruments, e.g.~IRIS spectroheliograms, SST, GST, Gregor, DKIST, and Solar-C.

 \textbf{Comparison between the properties of CJs and UV/EUV jetlets:}
Jetlets shoot out with an average speed of 70$\pm$30 \kms, which is similar to the average speeds of quiet-region (100$\pm$20 \kms) and coronal-hole (70$\pm$30 \kms) jets estimated by \cite{panesar16b,panesar18}. The average duration of jetlets is four times shorter (3$\pm$1m) than the average duration of CJs ($\sim$12 minutes \citealt{shimojo96,savcheva07,panesar16b,panesar18}). 
 Using AIA 171 \AA\ images (\iris\ Si IV SJI), we estimated that jetlet base width, spire length, and spire width have mean values of about 4400$\pm$2000 km (4400$\pm$2500 km), 27000$\pm$8000 km (16000$\pm$6000 km) and 3200$\pm$2000 km (3000$\pm$2500 km), respectively. The studied jetlets, on average, are at least three times smaller in base width ($<$5,000 km) than typical CJs ($\sim$18,000 km). 
 
\textbf{Comparison with spicules:} The observed jetlet speeds and durations are similar to Type-II spicule speeds (30-110 \kms) and durations (0.83-2.5 m), but their spire widths are six times larger than spicule widths (500 km; \citealt{sterling00b,pontieu07,pereira12}). Also, the occurrence rate of jetlets (some tens in 24 hrs in the Figure \ref{fig1}b FOV), is much less than that of spicules \citep[e.g.][]{beckers72}. 


Jetlets show some similarities with spicules: both features occur at the edges of network lanes, 
have similar velocities and durations, and can show twisting motions.  However, the coronal 
response of  jetlets is different from that of spicules.  So it could be that jetlets are on the 
smaller-size-scale end of features driven like CJs (cf.  \citealt{sterling16b}), while spicule 
might have a different driving mechanism with natural explanations for their 
velocities, durations, and twisting motions (e.g., \citealt{sykora17,iijima17}).

 
 Our recent CJ observations \citep{sterling15,panesar16b} apparently show (a) a pre-eruption minifilament sits at a PIL of a sheared bipole between a minority-polarity flux patch and majority-polarity flux patch; (b) due to continuous flux cancelation at the PIL, the minifilament field eventually becomes unstable and erupts outwards and a JBP appears at the PIL via \textit{internal reconnection}; (c) the outer envelope of the erupting minifilament field reconnects with the neighboring open field, which results in the formation of new open field lines and CJ material flowing out along these newly-opened lines. 

Our jetlet observations appear to be consistent with the CJ picture in some ways: Jetlets occur at PILs between merging minority and majority flux patches, the minority-polarity flux patch approaches (and cancels with) the majority-polarity network flux patch, and the flux cancelation plausibly prepares and triggers one or more small-scale flux rope eruptions that drive one or more jetlets. One of our jetlet spires (C3) seems to show twisting motion, which suggests that a highly twisted flux rope erupted from the jetlet-base. The erupting small-scale flux rope would result in internal reconnection in the erupting field, and brightenings/JBP (Figures \ref{fig2} and \ref{fig3}) appear at the eruption site. The spire would start to form when the erupting flux rope reconnects (\textit{external reconnection}) with encountered open field. The jetlet material would then escape along the newly-opened field lines. We do not however, see any clear signatures of brightenings of closed loops made by external reconnection  of the erupting field with the encountered open field as in the \cite{sterling15} CJ picture; we need further observations to determine whether this is due to the brightenings being too faint to observe with \iris\ and AIA, or because there is an inconsistency with the CJ picture.

\section{Conclusion} Our observations of ten jetlets suggest that flux cancelation is a necessary condition for the buildup and triggering of  UV/EUV network jetlets and they usually stem from the edges of  magnetic network flux lanes. Jetlets are therefore plausibly small-scale versions of the larger CJs, and of still-larger CMEs events. 
 
\vspace{0.9cm}

\newpage
\acknowledgments
N.K.P’s research was supported by an appointment to NPP at the  NASA/MSFC, administered by USRA under contract with NASA. A.C.S and R.L.M acknowledge the support from the NASA HGI program, and by the \Hinode\ Project. S.K.T acknowledges
support by NASA contract NNG09FA40C (IRIS). We acknowledge the use of \iris\ and \sdo\ data. B.D.P acknowledges support from NASA grants NNX16AG90G, and NNG09FA40C (IRIS). IRIS is a NASA small explorer mission developed and operated by LMSAL with mission operations executed at NASA Ames Research center and major contributions to downlink communications funded by ESA and the Norwegian Space Centre. 

\bibliographystyle{aasjournal}

\end{document}